\documentclass[10pt,conference,letterpaper]{IEEEtran}

\usepackage[utf8]{inputenc}
\usepackage{graphicx}
\usepackage{amssymb}
\usepackage[cmex10]{amsmath}
\usepackage{color}
\usepackage{theorem}
\usepackage{pifont}

\usepackage{bbm}

\renewcommand{\mod}{ \text{\, mod \,} }

\newcommand{\rpolx}[1]{\ensuremath{r^{(#1)}(x)}}
\newcommand{\rpol}[1]{\ensuremath{r^{(#1)}}}

\newcommand{\qpolx}[1]{\ensuremath{q^{(#1)}(x)}}
\newcommand{\qpol}[1]{\ensuremath{q^{(#1)}}}
\newcommand{\qhpolx}[1]{\ensuremath{\hat{q}^{(#1)}(x)}}

\newcommand{\qbpolx}[1]{\ensuremath{\bar{q}^{(#1)}(x)}}

\newcommand{\upolx}[1]{\ensuremath{u^{(#1)}(x)}}
\newcommand{\upol}[1]{\ensuremath{u^{(#1)}}}

\newcommand{\uhpolx}[1]{\ensuremath{\hat{u}^{(#1)}(x)}}

\newcommand{\apolx}[1]{\ensuremath{a^{(#1)}(x)}}
\newcommand{\appolx}[1]{\ensuremath{\bar{a}^{(#1)}(x)}}
\newcommand{\apol}[1]{\ensuremath{a^{(#1)}}}
\newcommand{\appol}[1]{\ensuremath{\bar{a}^{(#1)}}}
\newcommand{\DPx}[1]{\ensuremath{\Delta^{(#1)}(x)}}
\newcommand{\DPol}[1]{\ensuremath{\Delta^{(#1)}}}
\newcommand{\sval}[1]{\ensuremath{s^{(#1)}}}
\newcommand{\cval}[1]{\ensuremath{c^{(#1)}}}

\newcommand{\mfloor}[1]{\ensuremath{\left\lfloor #1 \right \rfloor}}
\newcommand{\dhalf}{\ensuremath{\mfloor{\frac{d-1}{2}}}}
\newcommand{\reff}[1]{(\ref{#1})}
\newcommand{\order}{\ensuremath{\mathcal O}}

\newtheorem{lemma}{Lemma}

\renewcommand{\deg}{\delta}
\newcommand{\degl}{\textnormal{deg}\ }

\IEEEoverridecommandlockouts 

\begin{document}

\title{A Fast Generalized Minimum Distance Decoder for Reed-Solomon Codes Based on the Extended Euclidean Algorithm}

\author{\IEEEauthorblockN{Sabine Kampf and Martin Bossert}
\IEEEauthorblockA{Institute of Telecommunications and Applied Information Theory\\
University of Ulm, Germany\\
\texttt{\{sabine.kampf | martin.bossert\}@uni-ulm.de}}
\thanks{This work was supported by the German Research Council "Deutsche Forschungsgemeinschaft" (DFG) under Grant No. Bo867/22.}}
\maketitle

\begin{abstract}
This paper presents a method to determine a set of basis polynomials from the extended Euclidean algorithm that allows Generalized Minimum Distance decoding of Reed-Solomon codes with a complexity of $\order(nd)$.
\end{abstract}


\section{Introduction}
Decoding of Reed-Solomon (RS) codes with the help of the extended Euclidean algorithm (EEA) was first presented by Sugiyama~et.~al. in 1975 \cite{Sugiyama}. In 1996, K\"otter introduced fast Generalized Minimum Distance (GMD) decoding of RS codes \cite{Koetter_GMD}. A first approach to combine GMD decoding and decoding with the EEA was presented in \cite{scc_wir}. 
However, the approach presented there does not allow decoding with a complexity less than $\order(d^3)$. 

GMD decoding consists mainly of two steps, the first is the calculation of a list of possible solution, and the second step is to choose one of the solutions from a list. In this paper, we investigate only the task of finding the list of solutions. The approach presented accomplishes this with complexity $\order(d^2)$. 

The paper is organized as follows: In the next section, we give the definition of RS codes and the polynomials used in decoding with the EEA. We also shortly recall the idea of GMD decoding. In Section \ref{sec:EEAmitsval}, we derive a new stopping criterion for the EEA and show how this can be used in the definition of the new basis polynomials. After the definition, we derive the amount of additional information necessary for decoding. We shortly recall the FIA in Section \ref{sec:FIA}, and show the modification that reduces the complexity. Section \ref{sec:concl} concludes the paper.

\section{Notations and Definitions} \label{sec:notations}
\subsection{RS Codes and Key Equation}
In this paper, an $\mathcal{RS} (n,k,d=n-k+1)$ code over $GF(q)$ with rate $R = \frac{k}{n}$ is defined in the spectral domain such that the spectrum of all codewords is zero at the first $n-k=d-1$ coefficients, hence 
\begin{equation}
C(x)=C_{d-1}x^{d-1}+\dots+C_{n-1}x^{n-1},
\label{eqn:cwdef}
\end{equation}
the information symbols $C_i \in GF(q),\;i={d-1},\dots,{n-1}$. The codeword $c(x)=c_0+c_1x+\dots+c_{n-1}x^{n-1}$ is calculated by the inverse discrete Fourier transform (IDFT):
\begin{equation}
c_i = n^{-1}\cdot C(\alpha^{-i}),\quad i=0,\dots,n-1,
\end{equation}
and conversely $C(x)$ can be recovered by applying the discrete Fourier transform (DFT) to $c(x)$:
\begin{equation}
C_j = c(\alpha^{j}), \qquad \, j=0,\dots,n-1.
\end{equation}
Thereby, let $\alpha \in GF(q)$ denote an element of order $n$. Throughout this paper, capital letters denote polynomials in the spectral domain, and small letters their correspondences in the time domain.

The transmitted codeword is corrupted by an additive error $e(x)$ of weight $t$, and the received word is $r(x)=c(x)+e(x)$. For decoding, calculate the syndrome $S(x)$:
\begin{equation}
S(x) = R(x) \mod x^{d-1} = E(x) \mod x^{d-1}.
\end{equation}
This syndrome is used in the \textit{key equation} for decoding RS codes:
\begin{equation} \label{eqn:key_eq}
-\Omega(x) \equiv \Lambda(x)\cdot S(x)  \mod x^{d-1},
\end{equation}
with the error locator polynomial $\Lambda(x)$ and the error evaluator polynomial $\Omega(x)$. These two polynomials satisfy the important degree relation:
\begin{equation}\label{eqn:deg_omega}
\degl\Omega(x) < \degl \Lambda(x) =t.
\end{equation}

\subsection{Decoding with the EEA}
Sugiyama~et.~al.~\cite{Sugiyama} showed that \reff{eqn:key_eq} can be solved using the EEA. The EEA uses the input polynomials $A(x)=\rpolx{0}$ and $B(x)=\rpolx{-1}$ to recursively calculate a series of quotient polynomials $\qpolx{j}$ and remainders $\rpolx{j}$ that fulfill:
\begin{equation}
\rpolx{j+1}=\rpolx{j-1} - \qpolx{j+1}\cdot\rpolx{j},
\label{eqn:EEAremrec}
\end{equation}
with $\degl \rpolx{j+1}<\degl \rpolx{j}$.
From the quotient polynomials, a series of auxiliary polynomials $\upolx{j}$ is obtained recursively, namely
\begin{equation}
\upolx{j+1}=\upolx{j-1} - \qpolx{j+1}\cdot\upolx{j},
\label{eqn:EEAauxrec}
\end{equation}
where $\upolx{-1}=0$ and $\upolx{0}=1$. The degrees of these polynomials are given by
\begin{equation}
\degl \upolx{j} = \sum_{i=1}^{j} \degl \qpolx{i}.
\end{equation}
Further, these polynomials fulfill the relation 
\begin{equation}
\upolx{j}\cdot A(x) = \rpolx{j} \mod B(x),
\end{equation}
which has a form similar to the key equation \reff{eqn:key_eq}. This implies that the EEA can be used for solving \reff{eqn:key_eq}.
Hence by setting $A(x) = S(x)$ and $B(x)=x^{d-1}$, in some steps of the EEA, whenever 
\begin{equation}
\degl \upolx{j} > \degl \rpolx{j}.
\label{eqn:EEAstopcond}
\end{equation}
we obtain polynomials fulfilling both \reff{eqn:key_eq} and \reff{eqn:deg_omega}.
If the number of errors $t$, i.e. the number of nonzero coefficients in $e(x)$, is limited by $t\leq \dhalf$, then it is known \cite{Sugiyama} that $t = \degl \upolx{j}$, $\Lambda(x) = \upolx{j}$ and $\Omega(x) = -\rpolx{j}$ if $j$ is the smallest index for which \reff{eqn:EEAstopcond} is fulfilled. 

Another property we will use is that \cite{Sugiyama}
\begin{equation}
\degl \upolx{j} + \degl \rpolx{j} \leq d-2.
\label{eqn:dm2eqn}
\end{equation}

Because our analysis relies strongly on the degrees of the polynomials, we introduce the abbreviation $\deg$ for the degree of a polynomial, i.e. $\deg\rpol{j} =\degl\rpolx{j}$, and equivalently for all other polynomials.

\subsection{GMD Decoding}
GMD decoding, introduced by Forney \cite{Forney_GMD}, is a method for soft-decision decoding by multi-trial decoding with a simple decoder. To accomplish this, a GMD decoder performs $m$ decoding trials. In each trial $j=1,\dots,m$, the $\tau_j$ least reliable symbols are erased. For GMD decoding, we take the polynomial $\Lambda(x)$ to be a joint error and erasure locator, so if the symbol at position $i$ is erased, we know that $\Lambda(\alpha^{-i})=0$. This means, that if we obtain $\Lambda(x)$ as a linear combination of polynomials \cite{scc_wir}
\begin{equation}
\Lambda(x) = \sum \beta_i \DPx{i},
\end{equation}
then each erasure gives us one equation for the determination of the coefficients $\beta_i$. If we find a proper locator polynomial, i.e. a polynomial of degree $t$ with exactly $t$ roots in GF(q), we store this polynomial in a list. After all trials have been performed, the GMD decoder selects one error locator which minimizes the error weight in a given metric.

For the description of the algorithm, it is not necessary to know the origin or calculation of the reliability information. Therefore, we assume that our decoder is provided with a list of positions, sorted by reliability. The positions are erased in order of reliability, with the least reliable position being erased first. Further, we do not address the problem of choosing a single solution in this paper. We only state that it is possible to solve this problem with quadratic complexity, too, in a way similar to the one presented in \cite{Koetter_GMD}. 

\section{A Closer Look at Decoding With the Extended Euclidean Algorithm}\label{sec:EEAmitsval}
\subsection{The Polynomials Calculated in the EEA}
As mentioned, if $t\leq \dhalf$, the error locator polynomial equals the polynomial $\upolx{j}$ of least index $j$ for which \reff{eqn:EEAstopcond} is fulfilled. We will now verify this limitation of the decoding radius in an unusual manner, thereby introducing a value $\sval{j}$ we will need later. 

Consider the syndrome polynomials $S(x)$. It was defined to be the known part of the spectrum, where the spectrum is assumed to be cyclically consecutive. This means that the (virtual) coefficient $S_{-1}$ is unknown. E.g. if the codeword is defined as in \reff{eqn:cwdef}, then $S_{-1}=E_{n-1}$ which is unknown because in general $C_{n-1} \neq 0$. Because $S(x)=\rpolx{0}$, we set $\sval{0}=-1$. Since $\rpolx{1}=\rpolx{-1}-\qpolx{1}\cdot \rpolx{0}$, we conclude that the unknown coefficient $S_{-1}$ now affects the virtual coefficient $\rpol{1}_{-1}$ and the coefficients $\rpol{1}_{0},\dots,\rpol{1}_{-1+\deg \qpol{1}}$ in $\rpolx{1}$ and these, too, become unknown. Therefore, we set $\sval{1}=\deg \qpol{1}-1$, to indicate the largest coefficient of $\rpolx{1}$ that is unknown. In the same way, we find for all iterations:
\begin{equation}
\sval{j} = \sval{j-1} + \deg\qpol{j} = \sum_{i=1}^j \deg \qpol{i} + \sval{0} = \deg \upol{j} - 1.
\label{eqn:svaldef}
\end{equation}
Of course we cannot use any of the unknown coefficients $\rpol{j}_{i}$, $i \leq \sval{j}$, in the determination of the next quotient polynomial $\qpolx{j+1}$. If \reff{eqn:EEAstopcond} is fulfilled we see that $\deg \rpol{j} \leq \sval{j}$ and we cannot proceed any further, since we do not know any element of the remainder $\rpolx{j}$. The following Lemma gives a more general statement. 

\begin{lemma}
In any step of the EEA, at most $\cval{j+1} \triangleq \deg \rpol{j}-\sval{j}$ coefficients of $\qpolx{j+1}$ can be calculated.
\end{lemma}
Due to the limited space, no proof is given here.

Next, we show that the number of coefficients that can be calculated limits the decoding radius to $\dhalf$. Namely, the decoder will only be able to correctly determine the auxiliary polynomial $\upolx{j}$ if $\deg \upol{j} \leq \dhalf$. In order to show this, recall that a formula similar to \reff{eqn:svaldef} exists for $\deg \rpol{j}$: From \reff{eqn:EEAremrec}, we see that
\begin{equation}
\deg \qpol{j+1} = \deg \rpol{j-1} - \deg \rpol{j},
\label{eqn:degqpolx}
\end{equation}
which we can rewrite to 
\begin{multline}
\deg \rpol{j} = \deg \rpol{j-1} - \deg \qpol{j+1} = \\
\deg \rpol{-1} - \sum_{i=1}^{j+1}\deg \qpol{i} = d-1 - \deg \upol{i+1}.
\label{eqn:degrrec}
\end{multline}

The following two lemmas show that the value $\cval{j+1}$ can also be used as a stopping criterion for the EEA. First, we show that if the classical decoding radius is exceeded, i.e. $\deg \upol{j} > \dhalf$, then we can never calculate a coefficient of the next quotient polynomial because $\cval{j+1}<0$. We do so by showing the complementary statement. Afterwards, we show that the next auxiliary polynomial can only be entirely calculated if with this, too, $\dhalf$ is not exceeded.
\begin{lemma}
If $\cval{j+1} \geq 0$ then $\deg \upol{j} \leq \dhalf$.
\end{lemma}
\begin{IEEEproof}
We use \reff{eqn:degrrec} and \reff{eqn:svaldef} to rewrite:
{\small
\begin{eqnarray}
\cval{j+1} = \deg \rpol{j}\hspace{-0.5ex} - \hspace{-0.5ex}\sval{j} &=& \deg \rpol{-1} -\hspace{-0.7ex} \sum_{i=1}^{j+1}\deg \qpol{i} - \sval{0} -\hspace{-0.7ex} \sum_{i=1}^j \deg \qpol{i} \label{eqn:degrmsre1}\\
&=&d-1+1 -2\sum_{i=1}^{j}\deg \qpol{i} - \deg \qpol{j+1}.
\label{eqn:degrmsre}
\end{eqnarray}}
Because $\deg \qpol{j+1} \geq 1$, we obtain
\begin{equation}
0 \leq \cval{j+1}\leq d-1 -2\deg \upol{j},
\end{equation}
which is equivalent to 
\begin{equation}
\deg \upol{j} \leq \frac{d-1}{2}.
\end{equation}
If $d$ is odd, $\dhalf = \frac{d-1}{2}$. If $d$ is even, then $\frac{d-1}{2}$ is not an integer, and $\deg \upol{j}$ cannot exceed $\dhalf$.
\end{IEEEproof}
Note, that if $\deg \upol{j} = \dhalf$, it is not possible to calculate the next quotient polynomial: If $d$ is odd, $\deg \rpol{j}\leq\frac{d-3}{2}$ due to \reff{eqn:dm2eqn}, which directly gives us $\cval{j+1}\leq 0$. If $d$ is even, then $\deg \rpol{j} \leq \frac{d-2}{2}$ and $\cval{j+1}\leq 1$. Since we always need to calculate $\deg \qpol{j+1}+1 \geq 2$ coefficients in the next quotient polynomial, we will not be able to calculate $\upolx{j+1}$ in this case.

The next lemma shows, that we are only able to determine the complete quotient polynomial $\qpolx{j+1}$ in the next iteration if $\deg \upol{j+1} \leq \dhalf$, i.e. we do not exceed the decoding radius in the next iteration. 
\begin{lemma}
\label{lem:degunext}
Let $\cval{j+1}>0$. The following relations hold:\\
\vspace*{0.3ex}
If $\cval{j+1} \geq \deg \rpol{j-1}-\deg \rpol{j}+1$, then $\deg \upol{j+1} \leq \dhalf$.\\
\vspace*{0.3ex}
If $\cval{j+1} < \deg \rpol{j-1}-\deg \rpol{j}+1$, then $\deg \upol{j+1} > \dhalf$.

\end{lemma}
\begin{IEEEproof}
We rewrite, using \reff{eqn:degrrec}:
\begin{eqnarray}
\deg \rpol{j-1}-\deg \rpol{j} &=& \deg \rpol{-1} - \deg \upol{j} - \deg \rpol{-1} +\deg \upol{j+1}  \\
&=&\deg \upol{j+1} - \deg \upol{j} 
\label{eqn:lem31}
\end{eqnarray}
Further, we can rewrite \reff{eqn:degrmsre1} to
$\cval{j+1} = d - \deg \upol{j} - \deg \upol{j+1}$.
Combining this with \reff{eqn:lem31}, we find that for the first case given in Lemma \ref{lem:degunext}:
\begin{eqnarray}
d - \deg \upol{j} - \deg \upol{j+1} &\geq& \deg \upol{j+1} - \deg \upol{j} + 1 \Leftrightarrow \nonumber \\
\deg \upol{j+1} &\leq& \frac{d-1}{2}. 
\label{eqn:smallerdhalf}
\end{eqnarray}
Because $\dhalf = \frac{d-1}{2}$ for odd values of $d$ and $\frac{d-1}{2}$  is not an integer if $d$ is even, it is possible to state
\begin{equation}
\deg \upol{j+1} \leq \dhalf.
\end{equation}

The second case is similar: 
\begin{eqnarray}
d - \deg \upol{j} - \deg \upol{j+1} &<& \deg \upol{j+1} - \deg \upol{j} + 1 \Leftrightarrow \nonumber \\
\deg \upol{j+1} &>& \frac{d-1}{2} \geq \dhalf. 
\label{eqn:biggerdhalf}
\end{eqnarray}
\end{IEEEproof}

\subsection{From the EEA to the Linear System of Equations}
Now we will derive the basis polynomials used for the FIA. Given that the syndrome polynomial is of sufficient degree, each error locator polynomial can be obtained as the normalized auxiliary polynomial $\uhpolx{i}$ in some step $i$ of the EEA. These auxiliary polynomials are calculated recursively, see \reff{eqn:EEAauxrec}.
We apply this recursion and find
\begin{eqnarray}
\hspace*{-3ex}\upolx{i} &=& \upolx{i-2} - \qpolx{i}\upolx{i-1} \label{eqn:recexp1}\\
&=& -\qpolx{i}\upolx{i-3} + \nonumber \\
&&\quad+(\qpolx{i}\qpolx{i-1}+1)\upolx{i-2}\\
&=& (\qpolx{i}\qpolx{i-1}+1)\upolx{i-4} -\nonumber \\
&&\quad - \qpolx{i}(\qpolx{i-1}\qpolx{i-2}+1)\upolx{i-3}\label{eqn:recexp2}\\
&=& \dots \nonumber
\end{eqnarray}
Hence, $\upolx{i}$ can always be obtained from any two polynomials $\upolx{i-i_0}$ and $\upolx{i-i_0-1}$ calculated during earlier steps of the EEA. 
Of course, the higher the degree of $\upolx{i-i_0}$ and $\upolx{i-i_0-1}$, the lower the degree of the polynomials that still have to be determined.

The proposed method therefore calculates two polynomials $\DPx{1}$ and $\DPx{2}$ from the EEA which are then multiplied by polynomials $\apolx{i}$ and $\appolx{i}$ respectively to obtain $\upolx{i}$, i.e.
\begin{equation}
\upolx{i} = \appolx{i}\DPx{1} + \apolx{i}\DPx{2}.
\end{equation}
If $t \leq \dhalf$, we can calculate $\apolx{i}$ and $\appolx{i}$ from the syndrome. But if $t > \dhalf$, it is necessary to use e.g. reliability information to fully determine $\apolx{i}$ and $\appolx{i}$. 
The intuitive solution is to choose $\DPx{1}$ and $\DPx{2}$ as two polynomials obtained from the EEA, while $\apolx{i}$ and $\appolx{i}$ are obtained by using a GMD decoding method. We first set
\begin{eqnarray}
\DPx{1} &=& \upolx{i_B} \quad \textnormal{and} \label{eqn:defdelta1} \\
\DPx{2} &=& \upolx{i_B-1},
\label{eqn:defdelta2}
\end{eqnarray}
where $i_B$ is such that $\deg \upol{i_B} \leq \dhalf$ and $\deg \upol{i_B+1} > \dhalf$. These are the polynomials of highest degree that are obtained from the EEA, leaving the determination of polynomials $\apolx{i}$ and $\appolx{i}$ of smallest degree.
When performing the recursive expansion as in \reff{eqn:recexp1} to \reff{eqn:recexp2} until $i-i_0=i_B$, then we find that $\appolx{i}$ consists of the sum of 
\begin{equation}
\prod_{j=i_B+1}^i\qpolx{j}
\end{equation}
and some terms where not all of the factors are present. The same holds for $\apolx{i}$ and $\prod_{j=i_B+2}^i\qpolx{j}$. Therefore, we find that
\begin{eqnarray}
\deg\apol{i} &=& \degl \prod_{j=i_B+2}^i\qpolx{j} \quad \textnormal{and} \label{eqn:apoldeg}\\
\deg\appol{i} &=& \degl \prod_{j=i_B+1}^i\qpolx{j},\label{eqn:appoldeg}
\end{eqnarray}
with the empty product being defined as $1$. Special care needs to be taken with $\deg \qpol{i_B+1}$ in case we stopped the EEA because $\deg \rpol{i_B} - \sval {i_B} < 0$: This condition implies, that the coefficient $\rpol{i_B}_{\sval{i_B}} = 0$. However, the definition of $\sval {i_B}$ tells us that this coefficient is unknown, hence we cannot be sure of $\deg \rpol{i_B}$. It is therefore reasonable to set
\begin{equation}
\deg \qpol{i_B+1} = \deg \rpol{i_B-1}-\max \left\lbrace\deg \rpol{i_B},\sval{i_B} \right\rbrace. 
\label{eqn:degqpolnew}
\end{equation}


On the other hand, we can do better if we stopped the EEA because $0 < \deg \rpol{i_B}-\sval{i_B} < \deg \rpol{i_B-1}-\deg \rpol{i_B}$. The first inequality tells us that we still can correctly determine some of the coefficients of $\qpolx{i_B+1}$, but the second inequality shows that we cannot determine the whole quotient polynomial. Denote the part of $\qpolx{i_B+1}$ with known coefficients as $\qhpolx{i_B+1}$, $\qbpolx{i_B+1}$ then is the part with unknown coefficients and $\qpolx{i_B+1} = \qhpolx{i_B+1} + \qbpolx{i_B+1}$. In this case, we define
\begin{eqnarray}
\DPx{1} &=& \upolx{i_B} \quad \textnormal{and} \\
\DPx{2} &=& \upolx{i_B-1} + \qhpolx{i_B}\cdot \upolx{i_B},
\end{eqnarray}
To see that this definition is reasonable, we write
\begin{multline}
\upolx{i_B+1} = \upolx{i_B-1} + \qhpolx{i_B+1}\upolx{i_B} \\+ \qbpolx{i_B+1}\upolx{i_B},
\end{multline}
i.e. $\deg \apolx{i_B+1} = 1$ and $\deg \appolx{i_B+1} = \deg \qbpolx{i_B+1}$. Compared to \reff{eqn:appoldeg}, we see that with this definition the number of unknown coefficients that need to be determined is smaller than before. For the next step, we find that 
\begin{multline}
\upolx{i_B+2}=
 \DPx{1}\cdot\left(1+\qpolx{i_B+2}\qbpolx{i_B+1}\right)\\ + \DPx{2}\cdot \qpolx{i_B+2}
\end{multline}
which is equivalent to \reff{eqn:apoldeg} and \reff{eqn:appoldeg} for the second step, only $\qpolx{i_B+1}$ now being replaced by $\qbpolx{i_B+1}$.
%
\subsection{Necessary Number of Erasures}
With the basis polynomials used in \cite{scc_wir}, we need $2t_0$ erasures if $t = \dhalf + t_0$ errors shall be corrected. We will now show that we need the same number of erasures for the proposed method.
First consider the situation as given in \reff{eqn:defdelta1} and \reff{eqn:defdelta2}, i.e. $\deg \rpol{i_B}-\sval{i_B}\leq 0$. The polynomial $\upolx{i_B+1}$ is the polynomial of least degree for that we need to apply GMD decoding. In order to determine this polynomial, we need to find the polynomials $\apolx{i_B+1}$ and $\appolx{i_B+1}$. Let $\deg \upol{i_B+1}=\dhalf+t_1$. According to \reff{eqn:degqpolnew}:
\begin{equation}
\begin{split}
\deg \appol{i_B+1} =&\deg \qpol{i_B+1}=\deg \upol{i_B+1}-\deg \upol{i_B} \\
=& \deg \rpol{i_B-1} - \sval{i_B}=\dots \\
=& d-2\deg \upol{i_B}.
\end{split}
\end{equation}
Combining the first and last row, one finds
\begin{equation}
\deg \upol{i_B} = d - \deg \upol{i_b+1} = \left \lceil \frac{d+1}{2} \right\rceil -t_1.
\end{equation}
For odd $d$ we find that $\deg \appol{i_B+1} = 2t_1-1$, while $\deg \apol{i_B+1}=0$, so the total number of unknown coefficients is 
\begin{equation}
\deg \appol{i_B+1}+1+\deg \apol{i_B}+1 = 2t_1+1.
\end{equation}
If $d$ is even, then $\deg \appol{i_B+1} = 2t_1-2$ and the total number of unknown coefficients is 
$2t_1$.
On the other hand, if the EEA was stopped because $0<\deg \rpol{i_B}-\sval{i_B} < \deg \rpol{i_B-1}-\deg\rpolx{i_B}$, we get
\begin{equation}
\begin{split}
\deg \appol{i_B+1}=& \deg \rpol{i_B-1}-\deg\rpolx{i_B}-\left(\deg \rpol{i_B}-\sval{i_B}\right) \\
=& \dots = -d+1+2\deg \upol{i_B+1}.
\end{split}
\end{equation}
From this we find that $\deg \appolx{i_B+1} = 2t_1-1$ if $d$ is odd, and $\deg \appolx{i_B+1} = 2t_1-2$ if $d$ is even, so we get the same total number of unknowns as before. 
Since one coefficient can always be chosen in order to normalize the error locator polynomial, we find that $2t_1$ erasures are enough to find $\deg \upol{i_B+1}$. For the further polynomials with $\deg \upol{i} = \dhalf + t_0$, $t_0>t_1$, we note that 
\begin{equation}
\deg \apol{i} = \degl \prod_{j=i_B+2}^i\qpolx{j} = \deg\appol{i}-\deg\appol{i_B+1}
\end{equation}
Thus, 
\begin{equation}
\deg\upol{i}=\deg\upol{i_B+1}+ \degl \prod_{j=i_B+2}^i\qpolx{j},
\end{equation}
and so we must have $\degl(\prod_{j=i_B+2}^i\qpolx{j})=t_0-t_1$. This directly yields the number of unknown coefficients
\begin{equation}
\deg \appol{i+1}+1+\deg \apol{i}+1 = 
2t_0+1,
\end{equation}
i.e. we need $2t_0$ erasures, because again one coefficient is chosen due to normalization. 

\section{The Fundamental Iterative Algorithm} \label{sec:FIA}
The original version of the FIA as introduced by Feng and Tzeng in \cite{Feng_Tzeng} gives the smallest set of linearly dependent leading columns of a matrix $\mathbf{A}$, together with the connection vector, indicating the vanishing linear combination. However, we again use the same modification as in \cite{scc_wir}, where we obtain all the solutions to all $2\tau \times (2\tau+1)$ submatrices that are situated in the upper left corner of $\mathbf{A}$. The FIA solves homogeneous, linear systems of equations, so we reformulate our problem. If $\deg \upolx{i_B+1}=\dhalf+1$, then instead of looking for two polynomials with $\deg \apol{i_B+1}=0$ and $\deg \appol{i_B+1}=1$, we search for a linear combination of the polynomials $\DPx{1}$, $x\cdot \DPx{1}$ and $\DPx{2}$; then additionally $x^2\cdot \DPx{1}$ and $x\cdot \DPx{2}$ when $\deg \apol{i_B+1}=1$ and $\deg \appol{i_B+1}=2$, and so on. It will be seen later that this choice allows us to decrease the complexity of the FIA to $\order(d^2)$.

We see in \reff{eqn:apoldeg} and \reff{eqn:appoldeg}, that sometimes $\deg \apol{j}$ and $\deg \appol{j}$ increase by more than one for the next step. We ignore this during the execution of the FIA. In such a case, the intermediate result should not give a valid error locator polynomial. But since the gap has the same size in both the sequence of degrees of $\apolx{j}$ and $\appolx{j}$, the next allowed solution will be obtained during one of the next steps of the FIA.

For a detailed description of the FIA, the reader is referred to \cite{Feng_Tzeng}. Here, we only note that the FIA starts the examination of each column with a connection vector $\boldsymbol{a}$, also called the starting vector. The FIA then calculates in each row a so-called discrepancy. If the discrepancy is zero, the connection vector is a valid solution for the current sub-system of equations, and the algorithm proceeds with the next row. If the discrepancy is non-zero, the connection vector is updated if possible, otherwise the vector and discrepancy are stored. 
The basic FIA has complexity $\order(d^3)$. It is known, cf. eg. \cite{Koetter_GMD}, that this complexity can be reduced if we can find a starting vector that allows us to save operations. We now show how this is possible with our basis polynomials.

The following matrix describes the system of equations that we want to solve with the FIA:
\begin{equation}\label{eqn:sysbetaMatrix}
\mathbf{A}=
\left(\begin{array}{c|c|c|cc}
\DPol{1}(\alpha_1)& \alpha_1\DPol{1}(\alpha_1) &  \DPol{2}(\alpha_1)& \alpha_1^2\DPol{1}(\alpha_1)&\cdots\\
\DPol{1}(\alpha_2)& \alpha_2\DPol{1}(\alpha_2) &  \DPol{2}(\alpha_2)& \alpha_2^2\DPol{1}(\alpha_2)&\cdots\\
\DPol{1}(\alpha_3)& \alpha_3\DPol{1}(\alpha_3) &  \DPol{2}(\alpha_3)& \alpha_3^2\DPol{1}(\alpha_3)&\cdots\\
\DPol{1}(\alpha_4)& \alpha_4\DPol{1}(\alpha_4) &  \DPol{2}(\alpha_4)& \alpha_4^2\DPol{1}(\alpha_4)&\cdots\\
\vdots & \vdots & \vdots &  \vdots 
\end{array}
\right).
\end{equation}
Assume that the vector $(a_1,a_2,a_3,\dots,a_{2i+1})$ solves the first $j$ equations of the $2i\times (2i+1)$ submatrix, i.e. the polynomial 
{\setlength{\arraycolsep}{0.5ex}
\begin{eqnarray}
\Lambda_1(x) &=& (a_1+a_2x+a_4x^2+\dots+a_{2_i}x^i)\cdot u_{t_B}(x) \nonumber \\
&&+ (a_3+a_5x+\dots+a_{2i+1}x^{i-1}) \cdot u_{t_B-1}(x)
\end{eqnarray}
}
has zeros for $\alpha_1, \dots, \alpha_j$. Then the vector $(0,a_1,0,a_2,a_3,\dots,a_{2i+1})$ of length $2i+3$ fulfills the first $j$ equations of the $(2i+2)\times (2i+3)$ subsystem of equations: Namely, this vector yields the polynomial 
\begin{multline}
(a_1x+a_2x^2+\dots+a_{2i}x^{i+1})\cdot u_{t_B}(x) \\
+ (a_3x+\dots+a_{2i+1}x^{i}) \cdot u_{t_B-1}(x)=x\cdot \Lambda_1(x).
\end{multline}
This polynomial has the same zeros as $\Lambda_1(x)$ plus an additional zero at $x=0$. Hence, by choosing this starting vector, it suffices to start the examination of the $(2i+3)$th column in row $j+1$. Due to the fact that in column $i$ we always take the connection vector stored in column $i-2$, it is necessary to store them separately for even and odd columns. Therefore we have to traverse the matrix from top to bottom twice, yet compared to the basic FIA where the matrix has to be traversed $\order(d)$ times, we are able to reduce the complexity to $\order(d^2)$.

\begin{figure}[htb]
\centering
\input{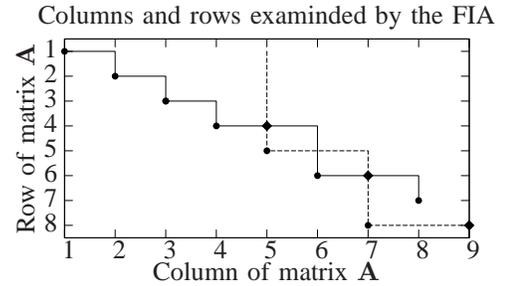}
\caption{Rows and columns examined by the FIA}
\label{fig:FIAexamcolskip}
\end{figure}

If $\deg \qpol{i_B+1}>1$, then $\deg \appol{i}-\deg \apol{i} = 2t_1-1$ and we need a slight modification to the algorithm described before. In order to keep the pattern of using the padded connection vector stored in column $i$ as starting vector in column $i+2$ in as many columns as possible, we write the evaluations of $\DPx{1}, x\DPx{1},\dots,x^{2t_1-1}\DPx{1}$ in the first $2t_1$ columns and $\DPx{2}$ in the $(2t_1+1)$th column. As starting vector for columns $i=2,\dots,2t_1$ we choose the connection vector stored in column $i-1$, padded with a zero in the first position. If the vector was stored for row $j$, we can start the examination of column $i$ in row $j$. In column $2t_1+1$, we start again in row 1, and for any future column $i$ use the connection vector stored in column $i-2$, padded with zeros in positions 1 and $2t_1+1$. Figure \ref{fig:FIAexamcolskip} shows the rows and columns examined by the FIA for $t_1 = 2$. The code used was an $\mathcal{RS} (16,6,11)$ over $GF(17)$. Here, all the points $(x,y)$ marked by a dot denote the point where a connection vector is stored, while the points marked with diamonds show at which point a connection vector was stored as a possible solution, i.e. in this case we obtain three candidate error locators. It can be seen, that the algorithm works very regular in columns $1$ through $6$. In column $7$ and row $7$, the discrepancy is zero, so that a vector is stored only in row $8$. This causes the third solution, stored in column $9$, not to include a term $ax^5\DPx{1}$.
Note, that this general case is consistent with the previous description for $t_1=1$.

\vspace*{-1ex}
\vspace*{-3ex}

\section{Conclusion}\label{sec:concl}
We presented a method to compute basic polynomials from the EEA that allow fast GMD decoding, because the list of possible solutions can be found with complexity $\order(d^2)$. Compared to \cite{scc_wir}, we gain one order of complexity, and achieve the same complexity as \cite{Koetter_GMD}. 
An approach to merge GMD decoding into the EEA, thereby superseding the use of the FIA, has been submitted to ITW 2010.

\section*{Acknowledgment}
The authors would like to thank Antonia Wachter and Vladimir Sidorenko for their valuable discussions and remarks. 

\bibliographystyle{IEEEtran}
\bibliography{bib_fast_gmd}

\end{document}